# Proportional Fairness in Multi-channel Multi-rate Wireless Networks – Part II: The Case of Time-Varying Channels with Application to OFDM Systems

Ying Jun (Angela) Zhang, *Member, IEEE* and Soung Chang Liew, *Senior Member, IEEE*

Department of Information Engineering, The Chinese University of Hong Kong
{soung, yjzhang}@ie.cuhk.edu.hk

***Abstract***- This is Part II of a two-part paper series that studies the use of the proportional fairness (PF) utility function as the basis for resource allocation and scheduling in multi-channel multi-rate wireless networks. The contributions of Part II are twofold. (i) First, we extend the problem formulation, theoretical results, and algorithms to the case of *time-varying* channels, where *opportunistic* resource allocation and scheduling can be exploited to improve system performance. We lay down the theoretical foundation for optimization that "couples" the time-varying characteristic of channels with the requirements of the underlying applications into one consideration. In particular, the extent to which *opportunistic optimization* is possible is not just a function of how fast the channel characteristics vary, but also a function of the elasticity of the underlying applications for delayed resource allocation. (ii) Second, building upon our theoretical framework and results, we study subcarrier allocation and scheduling in orthogonal frequency division multiplexing (OFDM) cellular wireless networks. We introduce the concept of a *W-normalized* Doppler frequency to capture the extent to which opportunistic scheduling can be exploited to achieve throughput-fairness performance gain. We show that a "look-back PF" scheduling can strike a good balance between system throughput and fairness while taking the underlying application requirements into account.

This work was supported under the Area of Excellence Scheme (Project Number AoE/E-01/99) and the Competitive Earmarked Research Grant (Project Number 414305 and Project Number 418506) established under the University Grant Committee of the Hong Kong Special Administrative Region, China.

# I. INTRODUCTION

Part I of this paper series has focused on the use of proportional fairness (PF) utility [1, 2] for resource allocation and scheduling in multi-channel multi-rate wireless networks. The problem formulation in Part I assumes the data rate enjoyed by a user *i* on a particular channel *k*, $b_{i,k}$, is constant and does not vary within the "application time window". Part II considers an alternative and complementary scenario where channel is fast-varying and the data rate enjoyed by a user on a channel does vary within the application time window. An application will be for an OFDM system, in which the data rates of users on different subcarriers may change with time due to fading, and such fast fading may be put to good use to boost system performance. With opportunistic scheduling, a subcarrier could be assigned to users who are enjoying good data rate at the moment [4], and by varying the allocated subcarrier airtimes to the users over time according to their momentary data rates, system performance can be improved.

### *Related Work and Connection with Part I*

The application domain of focus in Part II is OFDM wireless cellular systems. As mentioned in Part I, PF utility has recently attracted much attention in wireless cellular systems mainly because an opportunistic scheduler implemented in the High Data Rate (HDR) system [9] achieves PF bandwidth sharing among users under certain conditions [10, 11]. Specifically, the opportunistic scheduler serves one user at a time, and the user is selected according to the current data rate normalized by its average long-term throughput. By doing so, the logarithmic sum of the *long-run* average user throughputs is maximized. In [12], Kushner and Whiting analyzed the convergence of PF scheduling from a stochastic approximation's viewpoint. Instability of PF scheduling was observed by Andrews under non-saturated traffic condition [13]. Later, Borst proved that PF scheduling achieves stability whenever feasible when elastic traffic users have finite-size service demands [14]. Most of prior work is related to single-channel systems. In [15], Kim and Han extended the PF scheduler in the HDR system to multi-carrier systems. However, the characterization of the general PF utility in multi-channel wireless networks has not been well addressed to date.

The PF scheduler in the HDR system assumes that the underlying applications have high tolerance for delay. Consequently, it is the logarithmic sum of long-term average throughput that is of interest to most subsequent work on PF [10-15]. In practice, the elasticity of many elastic applications is limited. It is critical to guarantee that users receive a fair share of service within a *finite* application time window *W*. The scheduler has to deal with the fact that the channel state (data rate) of a user may not have visited every feasible state during *W*. Essentially, it is both the speed of channel variation and the length of the application time window that determine the extent to which opportunistic optimization is possible. More precisely, the utility should be written as $y = \sum_{i=1}^{U} \log(T_i^{(W)})$, where $T_i^{(W)}$ is the average throughput of user *i* within a time window *W*.

In our work here, we examine PF resource allocation in multi-channel multi-rate wireless networks where the "application time window" is taken into consideration in an explicit manner. The systems studied in [2'-7'] can be regarded as special cases of our system when $S = 1$ and/or $W \to \infty$. Part I of our work can be considered as the special case where the ratio of *W* versus the time scale at which user data rates change is small. For the WLAN motivating example in Part I, the user data rates $b_{i,k}$ can be rather static so that the value of *W* becomes immaterial. In Part II, we articulate that the result of Part I can still be relevant when $b_{i,k}$ is changing but *W* is small.

***Contributions***

The contributions and key results of Part II are summarized as follows:

- We extend the problem formulation, theoretical results, and algorithms to the case of *time-varying* channels, where *opportunistic* resource allocation and scheduling can be exploited to improve system performance. We lay down the theoretical foundation for optimization that "couples" the time-varying characteristic of channels with the requirements of the underlying applications into one consideration. In particular, the extent to which *opportunistic optimization* is possible is not just a function of how fast the channel characteristic varies, but also a function of the elasticity of the underlying applications for delayed resource allocation. In this paper, we express this elasticity through the concept of an application time window, *W*. Essentially, it is the throughput over *W* that will be looked at in the optimization process
- Building upon our theoretical framework and results, we study subcarrier allocation and scheduling in OFDM cellular wireless networks. We introduce the concept of a *W-normalized* Doppler frequency, which is the product of Doppler frequency [5] and *W*. The extent to which the scheduler can exploit opportunistic scheduling is tied to the *W*-normalized Doppler frequency and not just the physical Doppler frequency alone. We investigate several scheduling schemes optimizing different utilities, including maximizing system throughput, maximizing minimum user throughput, and maximizing PF with and without regard to the *W* of the underlying application. We show that a "look-back PF" scheduling scheme that takes into account $W$ can strike a good balance between system throughput and fairness while taking the underlying application requirements into account.

- For applications that have very high tolerance for delay, we can assume $W \to \infty$, for which we formulate a different approach that looks at the log utility of the "ensemble-averaged throughput". This formulation has two advantages from the implementation standpoint: (i) Its decision variables can be pre-computed off-line, so that real-time decision making consists of a mere table look-up; (ii) If for each user, the channel statistics of different channels are identically distributed (but not necessarily independently distributed), a reasonable assumption in many practical settings, the multi-channel optimization problem can then be reduced to the single-channel problem. This implies that the individual-channel PF optimality, which could be Pareto-inefficient under the general setting (see Part I), is Pareto-efficient and approximates joint-channel PF optimality when the application is highly delay-tolerant and the channel statistics are identically distributed on different channels. For the above reasons, our $W \to \infty$ formulation is a convenient low-complexity approximation when $W$ is large.

The remainder of this paper is organized as follows. Section II puts forth the problem formulation for the cases of $W = 1$, $1 < W < \infty$ , and $W \to \infty$. Section III discusses how the PF algorithms in Part I can be adapted for the random-rate case here, and presents several results on the characteristics of the PF solutions. Section IV uses the theory developed in Section III to investigate the performance of PF scheduling in cellular OFDM systems. Section V discusses how to extend the framework here for more general settings, and Section VI concludes this paper.

## II. Proportional Fairness under Time-Varying Random-Rate Channels and Application Time Window $W$

In Part I, the data rate enjoyed by user *i* on channel *k*, $b_{i,k}$, is deterministic and static. The airtime of channel *k* allocated to user *i*, $P_{i,k}$, is strictly based on the current $b_{i,k}$. No attempt is made to take into account the past throughput of user *i* as well as the possible future states of $b_{i,k}$. In Part II, in contrast, we consider the case where $b_{i,k}$ is random and changes dynamically. The variability of $b_{i,k}$ depends on the relative scales of two factors: (i) the rate at which $b_{i,k}$ changes with time; and (ii) the time scale of the underlying application requirement.

For applications that have very stringent real-time requirements, such as telephony, the approach in Part I is valid to the extent that throughput-fairness performance needs to be guaranteed within a small time window before $b_{i,k}$ can change appreciably. That is, under a bad channel condition, user *i* does not have the luxury to wait for $b_{i,k}$ to turn around. A decision has to be made based on the current $b_{i,k}$ because delay tolerance is small.

For non-real-time applications, such as large file download, what matters is not the throughput within the next few milliseconds, but rather the throughput over the next few seconds or even minutes. This situation is amenable to opportunistic scheduling and the throughput has to be viewed over a longer time scale.

In this paper, we consider the use of an application time window *W* to capture the time scale of the application's throughput requirement. Let $T_i[n]$ be the throughput of user *i* in time slot *n*. The quantity of interest to us is the "smoothed" throughput over a sliding window of *W*. Specifically, the normalized smoothed throughput in time slot *n* is

$$T_i^{(W)}[n] = \frac{1}{W} \sum_{m=n-n_0}^{n+W-n_0-1} T_i[m] \tag{1}$$

where $n_0$ is the number of slots we look back to the past, and $W - n_0 - 1$ is the number of slots we look to the future. Instead of using $T_i[n]$ as the throughput in the optimization problem, Part II uses the smoothed throughput $T_i^{(W)}[n]$.

Let us characterize $b_{i,k}$ as a stochastic process $B_{i,k}[n]$, $n = 0, 1, 2, ....$ , where we have adopted the convention that uppercase letters denote random variables and the corresponding lowercase letters denote specific realizations. The matrix $\mathbf{B}[n] = \left(B_{i,k}[n]\right)$ is the stochastic process corresponding to the whole system. When allocating airtimes in time slot *n*, we assume that the realizations of the channels in the current time slot *n* are known. That is, the matrix $\mathbf{b}[n] = \left(b_{i,k}[n]\right)$ is known. In addition, by time slot *n*, the decisions in the past have already been executed and therefore $T_i[n-1],\ T_i[n-2],\ ...$ in (1) are also known. The channel realizations in the future $\mathbf{b}[n+1],\ \mathbf{b}[n+2],\ ...$ are unknown during time slot *n*, and therefore $T_i[n+1],\ T_i[n+2],\ ...$ are also unknown, although they can be estimated based on the stochastic behavior of the channels. The utility function we will consider is [R1.5, R2.17]

$$E\left(\sum_i \log(T_i^{(W)}[n])\right) = E\left(\sum_i \log\left(\sum_{m=n-n_0}^{n+W-n_0-1} \frac{1}{W} T_i[m]\right)\right) = E\left(\sum_i \log\left(\sum_{m=n-n_0}^{n-1} \frac{1}{W} T_i[m] + \sum_{m=n}^{n+W-n_0-1} \frac{1}{W} T_i[m]\right)\right) \quad (2)$$

where the expectation is taken over $\mathbf{B}[n+1]$, $\mathbf{B}[n+2]$, ... (i.e., over the future evolution of the channels). Note that the first of the two sums on the right side of (2) consists of past throughputs that are known and cannot be changed anymore. The second of the two sums depend on the current and future decisions, as well as the future evolution of the channel states: $\mathbf{B}[n+1]$, $\mathbf{B}[n+2]$, ...

In this paper, where $W$ is finite, we will primarily look at the pure look-back scheme in which $n_0 = W-1$. We have chosen to focus on the look-back scheme because our numerical evaluation finds that the look-forward scheme does not yield better results and for many cases can be worse than the look-back scheme. In addition, the look-back scheme has the advantage that it does not have to predict the channel characteristics in the future, and can just rely on the past data to allocate resources. Perhaps an intuitive reason why the look-forward scheme does not yield better results is the uncertainty of channel realization in the future. With the pure look-back scheme, with reference to (2), our PF optimization problem in time slot $n$ is as follows:

$$\begin{aligned} &\max y[n] = E\left(\sum_i \log(T_i^{(W)}[n])\right) = \sum_{i=1}^{U} \log\left(a_i[n-1] + \frac{1}{W}\sum_{k=1}^{S} P_{i,k}[n] b_{i,k}[n]\right) \\ &\text{s.t.} \quad \sum_{i=1}^{U} P_{i,k}[n] = 1 \quad \forall\ k = 1,\cdots,S \\ &\qquad P_{i,k}[n] \ge 0 \qquad \forall\ i = 1,\cdots,U;\ k = 1,\cdots,S \end{aligned} \quad (3)$$

where $a_i[n-1] = \left(\sum_{m=n-W+1}^{n-1} T_i[m]\right) \big/ W$ for $n \ge W$ is obtained from the known throughputs in the past; $b_{i,k}[n]$ is the known bit rate of channel $k$ for user $i$ in the current slot; $P_{i.k}[n]$ is the decision variable corresponding to the fraction of the airtime of channel $k$ allocated to user $i$; $S$ is the number of channels; and $U$ is the number of users. In practice, the allocation of airtimes can be carried out in various ways in different systems. For example, in a TDMA system, a time slot could contain a frame consisting of multiple mini time slots. The fraction of airtime corresponds to the proportion of mini time slots allocated. In a random-access system, the fraction of airtime could be mapped to the aggressiveness of a user in its attempt to acquire the channel with respect to other users (e.g., the transmission probability used in ALOHA, or the contention window used in a back-off protocol). In this case, a time slot is defined as the period between two successive changes of the access parameters that regulate the aggressiveness of the stations [6].

For simplicity, let us suppose that the system start time is time slot 1. In addition, all users become active in time slot 1. Section V will discuss how to relax this assumption. A caveat in (2) is what to do for the initial time slots $n < W$, when we do not have enough time slots for smoothing purpose. For $n < W$, we could just smooth the throughput over *n* slots, and the optimization problem in (2) becomes

$$\begin{aligned} &\max y[n] = \sum_{i=1}^{U} \log\left( a_i[n-1] + \frac{1}{n}\sum_{k=1}^{S} P_{i,k}[n] b_{i,k}[n] \right) \\ &\text{s.t.} \ \sum_{i=1}^{U} P_{i,k}[n] = 1 \ \ \forall\ k = 1,\cdots,S \\ &\quad P_{i.k}[n] \ge 0 \qquad \forall\ i = 1,\cdots,U;\ k = 1,\cdots,S \end{aligned} \tag{4}$$

where $a_i[n-1] = \left(\sum_{m=1}^{n-1} T_i[m]\right)/n$. Note that the formulations in (3) and (4) apply to any window size *W*. In extreme cases $W = 1$ and $W \to \infty$, the formulation can be further simplified, as detailed below.

*$W = 1$ Case:*

Consider the $W = 1$ case. Then, $a_i[n-1] = 0$, and (2) becomes

$$\begin{aligned} &\max y[n] = \sum_{i=1}^{U} \log\left( \sum_{k=1}^{S} P_{i,k}[n] b_{i,k}[n] \right) \\ &\text{s.t.} \ \sum_{i=1}^{U} P_{i,k}[n] = 1 \ \ \forall\ k = 1,\cdots,S \\ &\quad P_{i.k}[n] \ge 0 \qquad \forall\ i = 1,\cdots,U;\ k = 1,\cdots,S \end{aligned} \tag{5}$$

which is of the same form as the problem formulation (1) in Part I. That is, the stringent application requirement of *W* = 1 deprives us of the chance to exploit time-varying channels for opportunistic scheduling, and the decision making process is the same as the deterministic case.

***$W \to \infty$ Case:***

The problem formulation for the asymptotic case $W \to \infty$ is slightly different. Again, for simplicity, let us suppose that the system start time is time slot 1. In addition, all users become active at time slot 1. Since $W \to \infty$, we will never be able to smooth the throughput over *W* as in (2). However, we could define the smoothed throughput at time *n* as $T_i^{(\infty)}[n] = \left(\sum_{m=1}^{n} T_i[m]\right)\Big/n$ so that the smoothing interval increases with *n*. Essentially, our problem formulation is that of (3) for all *n*.

We are faced with an implementation difficulty in the optimization problem (3). Namely, as *n* increases, the term $a_i[n-1]$ becomes more or less constant while the term $\left(\sum_{k=1}^{S} P_{i,k}[n] b_{i,k}[n]\right)\Big/n$ which contain the decision parameters $P_{i,k}[n]$ becomes smaller and smaller and vanishes to zero. This difficulty can be resolved in two ways: (i) impose a bound on $W$ that is sufficiently large; or (ii) where the statistical behaviors of the channels are known and ergodic, use ensemble average in the utility function rather than time average. For (i) the bound on $W$ should be sufficiently larger than the coherent times of the channels (the time over which $b_{i,k}[n]$ does not vary much [5]) so that $a_i[n-1]$ approaches a constant.

This paper will focus on approach (ii). As will be seen, it has the advantage that the solution can be pre-computed offline and stored so that during real time, the channel allocation decision consists of a mere table look-up. For (ii), the ensemble average of the throughput of user $i$ is the expectation of the per-time-slot throughput taken over all possible channel realizations of $\mathbf{B} = (B_{i,k})$, which we denote by $E_{\mathbf{B}}(T_i)$. We note that $E_{\mathbf{B}}(T_i) = \lim_{W\to\infty}\left(\sum_{m=1}^{W} T_i[m]\right)\big/ W$ for an ergodic system. The corresponding PF optimization can be phrased as

$$\begin{aligned}
&\max y = \sum_i \log(E_{\mathbf{B}}[T_i]) \\
&s.t.\ \ \sum_{i=1}^{U} P_{i,k|\mathbf{b}} = 1 \ \ \forall \ \ k = 1,...,S;\ \mathbf{b}\in\Omega \\
&\qquad P_{i,k|\mathbf{b}} \ge 0 \quad \forall \ \ i = 1,...,U;\ \mathbf{b}\in\Omega
\end{aligned} \tag{6}$$

where matrix $\mathbf{b} = \left(b_{i,k}\right)$ is a particular realization of $\mathbf{B}$ with $b_{i,k}$ being a particular realization of $B_{i,k}$, $\Omega$ is a set consisting of all possible realizations of $\mathbf{B}$, and $P_{i,k|\mathbf{b}}$ is the fraction of airtime on channel $k$ allocated to user $i$ given realization $\mathbf{b}$. Note that in this case, what matters is the "stationary" probability distribution of $\mathbf{B}$. Each realization of $\mathbf{B}$, $\mathbf{b}$, is mapped to a channel-airtime assignment represented by the matrix $\left( P_{i,k|\mathbf{b}}\right)$. The correlation between the channel states of successive time slots, $\mathbf{B}[n]$, $\mathbf{B}[n+1]$, $\mathbf{B}[n+2]$, ... , and the knowledge of the past throughputs $T_i[n-1]$, $T_i[n-2]$, ... , becomes immaterial.

Note that for the look-back formulation in (3) and (4), an optimization problem has to be solved in each and every time slot $n$ as $a_i[n-1]$ and realizations $b_{i,k}[n]$ become known. The ensemble average formulation in (6) for the $W\to\infty$ considers all possible realizations of $\mathbf{b} = \left(b_{i,k}\right)$ in one shot (to be elaborated in Subsection III.B). It has the advantage that the decision variables $P_{i,k|\mathbf{b}}$ are pre-computed off-line so that real-time decision making consists of a mere table lookup. Thus, even for finite but large *W*, it may be preferable to the use this formulation as an approximation.

Before leaving this section, we emphasize the following. For time-varying $b_{i,k}$, the problem formulation in (5) corresponds to one extreme in which the underlying application has very stringent time requirements, so that what matters to the application is the throughput it enjoys in each and every slot $T_i[n]$. It is identical to (1) in Part I, the deterministic $b_{i,k}$ case, even though the underlying assumption is that $b_{i,k}$ are time-varying. Problem formulations (3) and (4) concern applications that can tolerate throughput fluctuations within a time window $W > 1$, and what matters is the aggregate throughput within *W* time slots. Problem formulation (6) corresponds to the extreme case where $W \to \infty$ (or as an approximation for the large *W* case to exploit its advantage that the decision variables could be pre-computed offline). In this case, the user throughput to be looked at is the ensemble average of the per-slot throughput.

## III. Proportional-Fairness Algorithms and Characteristics of Solutions

The previous section has formulated the PF optimization problems. This section considers the solutions to the problems. The solutions to the *W*=1 case are the same as the deterministic case, and relevant discussions can be found in Part I. We focus on the $1 < W < \infty$ and $W \to \infty$ cases here.

### *A.* $1 < W < \infty$ *Case*

With respect to (3) and (4), define the vector $\mathbf{a}[n-1] = (a_i[n-1])$. Recall that we assume the channel realizations $\mathbf{b}[n] = (b_{i,k}[n])$, and $\mathbf{a}[n-1]$, are known in time slot *n* when we need to make the airtime allocation decision. To emphasize the dependency of the decision variable $P_{i,k}$ on $\mathbf{b}[n]$ and $\mathbf{a}[n-1]$, we will write $P_{i,k}[n]$ as $P_{i,k|\mathbf{b}[n],\mathbf{a}[n-1]}[n]$. The utility and shadow price are given by

$$y[n] = \sum_{i=1}^{U} \log\left( a_i[n-1] + \frac{1}{\min(n,W)} \sum_{k=1}^{S} P_{i,k|\mathbf{b}[n],\mathbf{a}[n-1]}[n] b_{i,k}[n] \right) \tag{7}$$

$$\frac{\partial y[n]}{\partial P_{i,k|\mathbf{b}[n],\mathbf{a}[n-1]}[n]} = \frac{b_{i,k}[n]}{\min(n,W) a_i[n-1] + \sum_{k=1}^{S} P_{i,k|\mathbf{b}[n],\mathbf{a}[n-1]}[n] b_{i,k}[n]} \tag{8}$$

The KKT conditions are similar to those in Part I, and the algorithms developed for Part I also apply here after replacement with the utility and shadow price here.

### *B.* $W \to \infty$

The optimization problem for the $W \to \infty$ case as formulated in (6) is more complicated than the $1 < W < \infty$ case. In the $1 < W < \infty$ case, the realizations $\mathbf{b}[n]$ and $\mathbf{a}[n-1]$ are known in time slot *n*, and based on the realizations, an optimization is performed in time slot *n*. In the $W \to \infty$ case, however, all the possible realizations of $\mathbf{B}$ must be considered a priori and in one shot. Implementation-wise, this means we do not perform optimization in each time slot *n* based on $\mathbf{b}[n]$ and $\mathbf{a}[n-1]$. Rather, the solution has to be pre-computed based on the statistical behavior of $\mathbf{B}$ and stored. In particular, a careful examination will reveal that the decision given one realization, say $\mathbf{B} = \mathbf{b}^{(1)}$, cannot be de-coupled from the decision given another realization $\mathbf{B} = \mathbf{b}^{(2)}$, because $P_{i,k|\mathbf{b}^{(1)}}$ and $P_{i,k|\mathbf{b}^{(2)}}$ cannot be independently determined based only on $\mathbf{b}^{(1)}$ and $\mathbf{b}^{(2)}$, respectively. Nevertheless, we will show in this section that, the general principles of the PF algorithms developed in Part I are still applicable for the $W \to \infty$ case after a problem transformation in which each channel realization $\mathbf{b}$ is mapped to a *virtual* channel.

The roadmap of our presentation on the $W \to \infty$ case is as follows. We start with the simple *U*-user-1-channel case in B.1, the result of which will be used in the general *U*-user-*S*-channel case later. In subsection B.1.1, the *U*-user-1-channel case is shown to be equivalent to the deterministic-channel *U*-user-multiple-channel case investigated in Part I after a “virtual channel mapping”. In subsection B.1.2, we show that the problem can be solved with much lower computational complexity when data rate $b_{i,k}$’s are continuous random variables. Subsection B.2 begins treating the general *U*-user-*S*-channel case by first arguing that the state space of the problem could grow exponentially and become intractable. Fortunately, the state space could be greatly reduced under two conditions: (i) when $B_{i,k}$ for different *k* are identically distributed (i.d.); and (ii) when $B_{i,k}$ for different *i* are independent identically-distributed (i.i.d.). In particular, the *U*-user-*S*-channel case is equivalent to the *U*-user-1-channel case under condition (i), and hence algorithms developed in subsection B.1 can be directly applied.

*B.1 U*-user-1-channel

Let us label the sole channel as channel *k*. We need to decide the airtime assignments to the *U* users, $P_{i,k|\mathbf{b}}$ for $i = 1, 2, ...U$, under different channel realizations $\mathbf{b}$, $\mathbf{b} \in \Omega$ here. The formulation in (6) becomes

$$\begin{aligned} &\max y = \sum_i \log(E_{\mathbf{B}}[T_i]) \\ &s.t.\ \sum_i P_{i,k|\mathbf{b}} = 1 \ \ \forall\ \mathbf{b} \in \Omega \\ &\quad P_{i,k|\mathbf{b}} \geq 0 \quad \forall\ i = 1,...,U;\ \mathbf{b} \in \Omega \end{aligned} \tag{9}$$

*B.1.1 PF optimization for single-channel systems with discrete* $b_{i,k}$

Here we assume $b_{i,k}$ is a discrete random variable first, and will consider continuous $b_{i,k}$ later. We assume that there are *M* possible bit rates. That is, $b_{i,k}$ has *M* possible realizations. Note that here we do not presume the different channels to be statistically identical: the set of possible discrete bit rates of different channels are the same, but the probabilities of their realizations are different.

Considering the channels of all *U* users, there are altogether $|\Omega| = M^U$ possible realizations of **B**. Denote these realizations by $\mathbf{b}^{(m)} = (b_{i,k}^{(m)}),\ m = 1,...,M^U$ . Define $p_{\mathbf{B}}\left(\mathbf{b}^{(m)}\right) = \Pr[\mathbf{B} = \mathbf{b}^{(m)}]$ (i.e., the probability of a particular realization). Then,

$$E_{\mathbf{B}}[T_i] = \sum_{\mathbf{b}^{(m)} \in \Omega} P_{i,k|\mathbf{b}^{(m)}} b_{i,k}^{(m)} p_{\mathbf{B}}\left(\mathbf{b}^{(m)}\right) \quad \text{for} \quad i = 1,\ ...,\ U. \tag{9}$$

It can be seen by direct comparison that the *U*-user-1-channel case above is equivalent to the deterministic-channel *U*-user-$M^U$-channel case in Part I with the following mapping. Each realization $\mathbf{b}^{(m)}$ corresponds to one "virtual channel", which we will label by $(k, \mathbf{b}^{(m)})$. On virtual channel $(k, \mathbf{b}^{(m)})$, the virtual bit rate of user *i* is $r_{i,(k,\mathbf{b}^{(m)})} = b_{i,k}^{(m)} p_{\mathbf{B}}\left(\mathbf{b}^{(m)}\right)$. This means the algorithms developed in Part I can be reused here, although the complexity is high. To limit scope, this paper will not delve too deeply into the complexity issue and will not attempt to identify fast heuristic, but suboptimal, algorithms. We are mainly interested in the characteristics of the PF solutions so that we can compare them with the solutions that adopt other utility functions. So, we only need algorithms that are fast enough for that purpose.

*B.1.2 PF optimization for single-channel systems with continuous* $b_{i,k}$

The state space of the computation for the discrete-rate case grows quickly with the number of possible values that can be adopted by $b_{i,k}$. At first glance, this implies a formidable problem for the continuous-rate case, since the number of possible values for $b_{i,k}$ is infinite. In contrast to (10), the expected throughput of user *i* for the continuous case is calculated as

$$E_{\mathbf{B}}[T_i] = \int_{\mathbf{b}\in\Omega} P_{i,k|\mathbf{b}} b_{i,k} f_{\mathbf{B}}(\mathbf{b}) d\mathbf{b} \tag{11}$$

where $f_{\mathbf{B}}(\mathbf{b})$ is the probability density function of $\mathbf{B}$ at $\mathbf{b}$.

We now show that the problem can actually be solved with a much lower computational complexity by making use of the fact that the probability that a channel is shared by more than one user is negligible in the continuous-rate case. For easy understanding, we first consider the 2-user-1-channel case before moving to the general *U*-user-1-channel case.

*2-user-1-channel case*

Recall that in Part I of this paper series where deterministic $b_{i,k}$ is assumed, we can sort $b_{1,k}/b_{2,k}$ from large to small for the solution of the 2-user-*S*-channel case. There is a boundary channel $S^*$ such that for all $k < S^*$ (where $b_{1,k}/b_{2,k} > T_1^*/T_2^*$), the channel will be exclusively used by user 1, and for all $k > S^*$ (where $b_{1,k}/b_{2,k} < T_1^*/T_2^*$), the channel will be exclusively used by user 2. The boundary channel $S^*$ may be shared by the 2 users, if $b_{1,S^*}/b_{2,S^*} = T_1^*/T_2^*$ (see Section IV.A in Part I).

Consider the infinite-*W* 2-user-1-channel case here. For the single physical channel, there are many virtual channels, each corresponding to a pair of bit-rate realizations of the two users. In the continuous-rate case, there are an infinite number of virtual channels $(k,\mathbf{b})$, because the number of possible realizations $\mathbf{b}$ is infinite. We note, however, that no matter how large the number of virtual channels is, there is still a boundary virtual channel that divides exclusive assignments of virtual channels to the two users.

Conceptually, to find the boundary virtual channel, we could sort the virtual channels $(k,\mathbf{b})$ according to the bit-rate ratio

$$b_{1,k} f_{\mathbf{B}}(\mathbf{b}) d\mathbf{b} \big/ b_{2,k} f_{\mathbf{B}}(\mathbf{b}) d\mathbf{b} \tag{11}$$

from large to small. Note that $f_{\mathbf{B}}(\mathbf{b}) d\mathbf{b}$ gets cancelled out in the numerator and denominator. Thus, we are still sorting according to $b_{1,k}/b_{2,k}$. The KKT condition [7] states that if $b_{1,k}/b_{2,k} > E_{\mathbf{B}}[T_1^*]/E_{\mathbf{B}}[T_2^*]$, then the channel is exclusively allocated to user 1; otherwise the channel is exclusively allocated to user 2. Since $B_{i,k}$ is a continuous random variable, the probability that $b_{1,k}/b_{2,k}$ is exactly equal to $E_{\mathbf{B}}[T_1^*]/E_{\mathbf{B}}[T_2^*]$ is zero. Hence, we can ignore the case where the channel is shared by user 1 and user 2. As a result, the PF optimization problem translates to the following fixed point problem:

$$\begin{aligned} E_{\mathbf{B}}[T_1^*] &= \int_{b_{1,k}} \int_{b_{2,k} \le b_{1,k} E[T_2^*]/E[T_1^*]} b_{1,k} f_{B_{1,k}}\left(b_{1,k}\right) f_{B_{2,k}}\left(b_{2,k}\right) db_{2,k} db_{1,k} \\ &= \int_{b_{1,k}} b_{1,k} f_{B_{1,k}}\left(b_{1,k}\right) F_{B_{2,k}}(b_{1,k} E[T_2^*]/E[T_1^*]) db_{1,k} \end{aligned}, \quad (13)$$

$$\begin{aligned} E_{\mathbf{B}}[T_2^*] &= \int_{b_{2,k}} \int_{b_{1,k} \le b_{2,k} E[T_1^*]/E[T_2^*]} b_{2,k} f_{B_{2,k}}\left(b_{2,k}\right) f_{B_{1,k}}\left(b_{1,k}\right) db_{1,k} db_{2,k} \\ &= \int_{b_{2,k}} b_{2,k} f_{B_{2,k}}\left(b_{2,k}\right) F_{B_{1,k}}(b_{2,k} E[T_1^*]/E[T_2^*]) db_{2,k} \end{aligned} \quad (14)$$

where we have assumed that the bit rates of user 1 and user 2 are independent of each other. That is,

$$f_{\mathbf{B}}\left(\mathbf{b}\right) = f_{B_{1,k}}\left(b_{1,k}\right) f_{B_{2,k}}\left(b_{2,k}\right). \quad (15)$$

This is a reasonable assumption for most communications systems. Likewise, $F_{B_{1,k}}(b_{1,k})$ and $F_{B_{2,k}}(b_{2,k})$ denote the cumulative distribution function (CDF) of $B_{1,k}$ and $B_{2,k}$, respectively. Having solved (13)-14) and using standard fixed-point algorithms, we can then allocate the virtual channel $(k,\mathbf{b})$ to whichever user having the larger $b_{i,k} / E_{\mathbf{B}}[T_i^*]$. That is, if the realization in time slot *n* is $\mathbf{b} = \left(b_{i,k}\right)$, then the user with the largest $b_{i,k} / E_{\mathbf{B}}[T_i^*]$ uses the physical channel exclusively.

*U-user-1-channel case*

The argument for the $U > 2$ case is largely the same. As there are an infinite number of virtual channels, we can again ignore the probability of sharing the channel between more than one user. By considering the KKT condition, the virtual channel $(k,\mathbf{b})$ should be assigned to the user with the largest virtual rate-throughput ratio $r_{i,(k,\mathbf{b}^{(m)})} / E_{\mathbf{B}}[T_i^*] = b_{i,k} f_{\mathbf{B}}\left(b\right) d\mathbf{b} / E[T_i^*]$. Since the term $f_{\mathbf{B}}\left(b\right) d\mathbf{b}$ is common to all users, this means the user with the largest $b_{i,k} / E_{\mathbf{B}}[T_i^*]$ will get the virtual channel. Thus, given channel realization $\mathbf{b}$, the solution that satisfies the following is optimal:

$$i^* = \arg\max b_{i,k} / E_{\mathbf{B}}[T_i^*] \quad (14)$$

$$P_{i^*,k|\mathbf{b}} = 1, \;\; P_{i,k|\mathbf{b}} = 0 \;\; \forall i \ne i^* \quad (15)$$

where $E_{\mathbf{B}}[T_i^*]$ can be obtained by solving the following fixed-point problem:

$$E_{\mathbf{B}}[T_i^*] = \int_{b_{i,k}} b_{i,k} f_{B_{i,k}}\left(b_{i,k}\right) \prod_{j \ne i} F_{B_{j,k}}\left(b_{i,k} E_{\mathbf{B}}[T_j^*] / E_{\mathbf{B}}[T_i^*]\right) db_{i,k} \quad \forall \; i \quad (16)$$

*B.2 U-user-S-channel*

Having discussed the PF algorithms for single-channel systems, we now move to systems with multiple physical channels. In the discrete-rate case, the random *U*-user-*S*-channel can be mapped to a deterministic *U*-user-$SM^{SU}$-channel case. To see this, with reference to the first constraint in (6), note that on each of the *S* physical channels, there are $|\Omega| = M^{SU}$ realizations of *B* to consider, since there are *SU* entries in the matrix *B* and each entry has *M* possible values. The state space for the optimization can be rather large indeed. In contrast, the computational complexity for the continuous-rate case is not much higher than that in single-channel systems, as shown in the following.

For the *U*-user-*S*-channel case, *k* becomes a channel index rather than the label of a particular physical channel. Solution characterizations (16) and (17) remain valid with $E_{\mathbf{B}}[T_i^*]$ in (18) being augmented by a summation over the *S* physical channels. Specifically, (16)-(18) are replaced by the following:

Given a channel realization $\mathbf{b} = (b_{i,k})$, for each physical channel $k = 1,...,S$,

$$i^*(k) = \arg\max b_{i,k} \big/ E_{\mathbf{B}}[T_i^*]; \tag{19}$$

$$P_{i^*(k),k|\mathbf{b}} = 1, \;\; P_{i,k|\mathbf{b}} = 0 \;\; \forall i \neq i^*(k); \tag{20}$$

where $E_{\mathbf{B}}[T_i^*]$ is solved by the fixed point problem below:

$$E_{\mathbf{B}}[T_i^*] = \sum_{k=1}^{S} \int_{b_{i,k}} b_{i,k}\, f_{b_{i,k}}(b_{i,k}) \prod_{j \neq i} F_{B_{j,k}}\left(b_{i,k}\, E_{\mathbf{B}}[T_j^*] \big/ E_{\mathbf{B}}[T_i^*]\right) db_{i,k} \quad \forall\, i \tag{21}$$

*State-Space Reduction*

In this subsection, we will show that the state space of the *U*-user-*S*-channel case can be reduced under two situations: (i) when $B_{i,k}$ for different *k* are i.d.; and (ii) when $B_{i,k}$ for different *i* are i.i.d., as detailed below.

*Case (i) For each i, $B_{i,k}$ for different k are i.d. random variables*

Suppose that for any given user *i*, the random variables $B_{i,k}$ for different physical channels *k* are identically distributed. For the subcarrier allocation problem, for example, this is a reasonable assumption provided the frequency separations between the different subcarriers are not very large (so that the reflection, refraction, and diffusion behaviors of the EM waves are almost the same). Note that with this assumption, at any particular time instant, the realizations $b_{i,k}$ for different *k* may not be equal due to different fadings on different channels. Over the long term, however, the probability distributions of $B_{i,k}$ are identical if we assume the different channels experience the same path loss and shadowing.

***Theorem 1:*** Suppose that for each user *i*, the random variables $B_{i,k}$ for different channels *k* are identically distributed. An optimal solution to the joint-channel optimization problem (6) is given by solving the single-channel optimization problem (9) and then applying the airtime mapping $P_{i,k|\mathbf{b}}$ in the similar way to all the *S* channels.

***Comment:*** Note that $B_{i,k}$ for different *i* need not be i.d. here. Also, for a user *i*, $B_{i,k}$ and $B_{i,l}$ of two different channels *k* and *l* need not be uncorrelated or independent.

Since solving the single-channel problem yields the solution to the joint-channel problem, the complexity in this case is the same as that of (9).

***Proof:*** See Appendix I.

*Case (ii) For each k,* $B_{i,k}$ *for different i are i.i.d. random variables*

Suppose that for any given physical channel *k*, the random variables $B_{i,k}$ for different users *i* are i.i.d.. This corresponds to the situation where long-term power control is applied so that the average received powers for all users over the long term are equal, and where the instantaneous received powers of users may vary due to fading.

**Theorem 2:** Suppose that for each channel *k*, the random variables $B_{i,k}$ for different users are i.i.d. An optimal solution to the joint-channel optimization problem (6) is given by choosing the user(s) with the maximum realization of $B_{i,k}$ to transmit on each and every channel *k* at any given time instant. Specifically, given a channel realization **b**, for each channel *k*, $P_{i,k|\mathbf{b}} = 1/w_k$ if $b_{i,k} = \max_j b_{j,k}$, and $P_{i,k|\mathbf{b}} = 0$ otherwise; where $w_k$ is the number of users *i* with $b_{i,k} = \max_j b_{j,k}$.

**Comment:** The solution above corresponds to the *deterministic-channel* case in which the utility function to be maximized is the system throughput.

***Proof:*** See Appendix I.

In this case, the solution is actually quite trivial and can be computed during time slot *n* rather than a priori. For each channel *k*, we need to identify the maximum(s) among *U* variables: $B_{i,k}$ for $i = 1, 2, ..., U$. So, given a particular realization, for the *S* channels, the computation required to make the airtime assignment decision is of order $O(SU)$. It is interesting to note that in this case, the problem of optimizing $\sum_i \log E_{\mathbf{B}}[T_i]$ here is actually simpler computationally than optimizing $\sum_i \log T_i$ in Part I.

It is also easy to show that PF scheduling ( $\max \sum_i \log E_{\mathbf{B}}[T_i]$ ), max-min scheduling ($\max \min \sum_i E_{\mathbf{B}}[T_i]$), and maximum-throughput scheduling ($\max \sum_i E_{\mathbf{B}}[T_i]$) yield the same result in this case.

## IV. NUMERICAL RESULTS: SUBCARRIER ASSIGNMENT PROBLEM IN CELLULAR NETWORKS

To illustrate the application of the theory as set up in the preceding section, this section moves on to the subcarrier assignment problem in cellular OFDM systems. In the simulations, an OFDM system with 16 subcarriers and symbol duration of 4 $\mu s$ is considered. The channel is assumed to be frequency-selective Rayleigh fading with an exponential delay profile, and the Doppler spread is 30Hz. We assume that there are four users in the system. For each of the following experiments, we run 100 independent replications of simulations, with each one lasting for 1 second. While the channel evolves as a correlated random process during each of the 1-second simulations, the channel realizations in different replications are independent of each other.

For the numerical studies, instead of looking at the discrete $b_{i,k}$, we assume continuous $b_{i,k}$ that is related to signal-to-noise (SNR) by the Shannon's Capacity formula: $b_{i,k} = \log_2(1 + SNR_{i,k})$, where $SNR_{i,k}$ is the received SNR of user *i* on sub-channel *k*. The main reason for considering continuous $b_{i,k}$ as such is that we are more interested in fundamental results rather than results due to the set of discrete rates imposed by a specific system design.

We investigate the performance of subcarrier assignment schemes where the length of the application time window, *W*, varies depending on the nature of the underlying application. Specifically, we study (1) PF scheduling for time-varying channels (i.e., the scheme developed in Section III.A, hereafter referred to as look-back PF); (2) PF scheduling assuming *W*=1 (i.e., the scheme in Part I, hereafter referred to as *W*=1 PF); (3) PF scheduling assuming $W \to \infty$ (i.e., the scheme developed in Section III.B, hereafter referred to as infinite-*W* PF); (4) maximum throughput (MT) scheduling; and (5) max-min fair scheduling. We note that max-min fair scheduling schemes for *W*=1 and $1 < W < \infty$ yield the same result, if a look-back method is adopted.

In the first set of experiment, we assume that the channels of all users are statistically identical, with a mean SNR of 13dB. RMS delay spread of the channel is equal to 216.5*ns*. In Fig. 1, we compare the throughput performance of look-back PF with other schemes. The figure shows that infinite-*W* PF and MT coincide with each other, as $B_{i,k}$ for different users are identical in this experiment. This result is consistent with Theorem 2. It is not surprising to see that when normalized Doppler frequency is small, the throughput of look-back PF is similar to that of *W*=1 PF, because $b_{i,k}$ hardly varies within an application time window. When normalized Doppler frequency increases, the throughput of look-back PF also increases, thanks to its ability to exploit time-domain diversity. When *W* becomes very large, convergence of look-back PF, infinite-*W* PF and MT is observed. The throughputs of max-min and *W*=1 PF, on the other hand, fail to make use of the diversity in time. Although MT and infinite-*W* PF achieve the highest throughputs, they try to exploit diversity in time too aggressively without giving due consideration to the actual *W* dictated by the specific application requirement. This causes them to perform poorly in terms of fairness when *W* is not that large.

The fairness performance is investigated in Fig. 2. In particular, Jain's fairness index [8] defined as

$$\frac{1}{N}\sum_{n=1}^{N}\left(\left|\sum_{i=1}^{U}T_i^{(W)}[n]\right|^2 \Big/ U\sum_{i=1}^{U}\left(T_i^{(W)}[n]\right)^2\right) \tag{22}$$

is plotted against normalized Doppler frequency. From the figure, it can be seen that max-min scheduling yields the maximum fairness regardless of the window size. On the other hand, MT scheduling maximizes system throughput at the cost of poor fairness when normalized Doppler frequency is small. When the application time window becomes large, users eventually get similar services as the channel statistics are identical. Therefore, the fairness indices of all schemes converge to 1. A close observation of Fig. 1 and Fig. 2 indicates that look-back PF strikes a good balance between throughput and fairness.

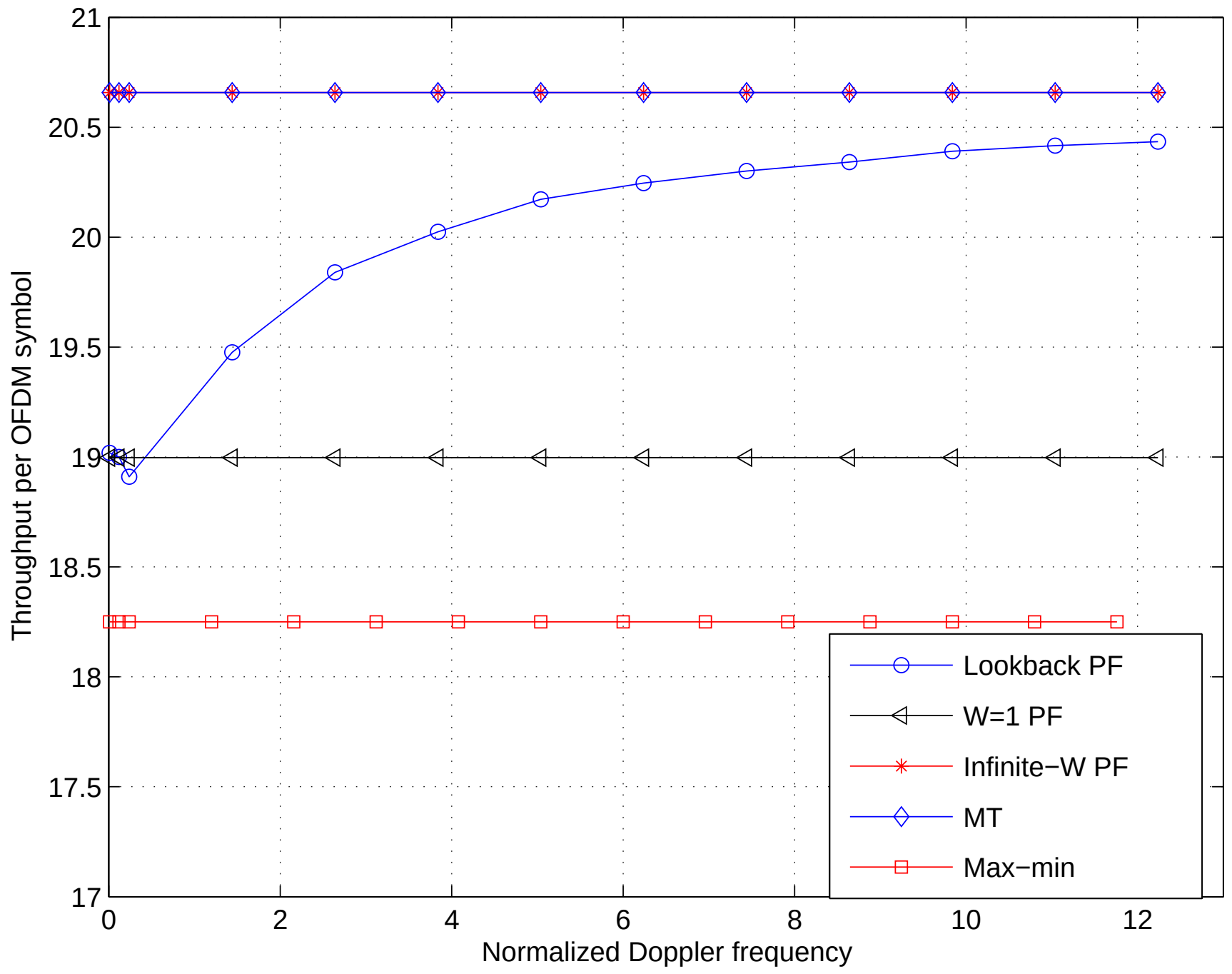

**Fig. 1:** System throughput as a function of normalized Doppler frequency

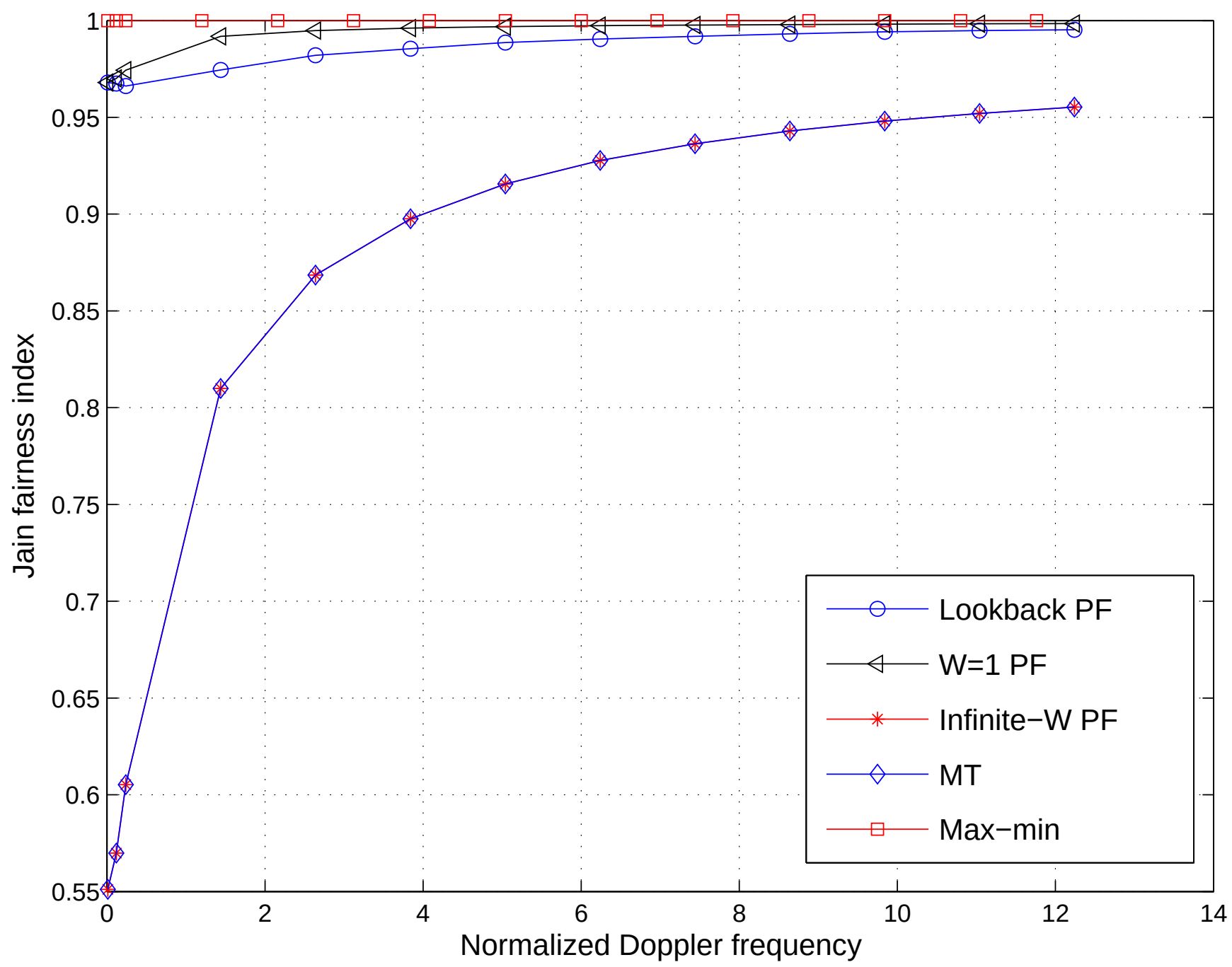

**Fig. 2:** Jain's fairness index as a function of normalized Doppler frequency

We now move to the case in which the channel statistics of different users are not identical. In Fig. 3 and Fig. 4, we assume that the mean SNR of the four users are 10dB, 12dB, 14dB, and 16dB, respectively. Similar to the homogeneous-user case, MT and max-min are optimized for two extremes, i.e., maximum throughput and maximum fairness, respectively. Interestingly, MT no longer coincides with infinite-*W* PF and its fairness index is much lower than the other schemes even with large normalized Doppler frequency. This is due to the fact that channel statistics are no longer identical, and MT always favors the user with the largest channel gain. In contrast, look-back PF automatically takes into account *W* when exploiting time-domain diversity to achieve a good tradeoff between throughput and fairness. Similar to the homogeneous-user case, the performance of look-back PF coincides with *W*=1 PF when the time window is small and converges to infinite-*W* PF when the time window becomes very large.

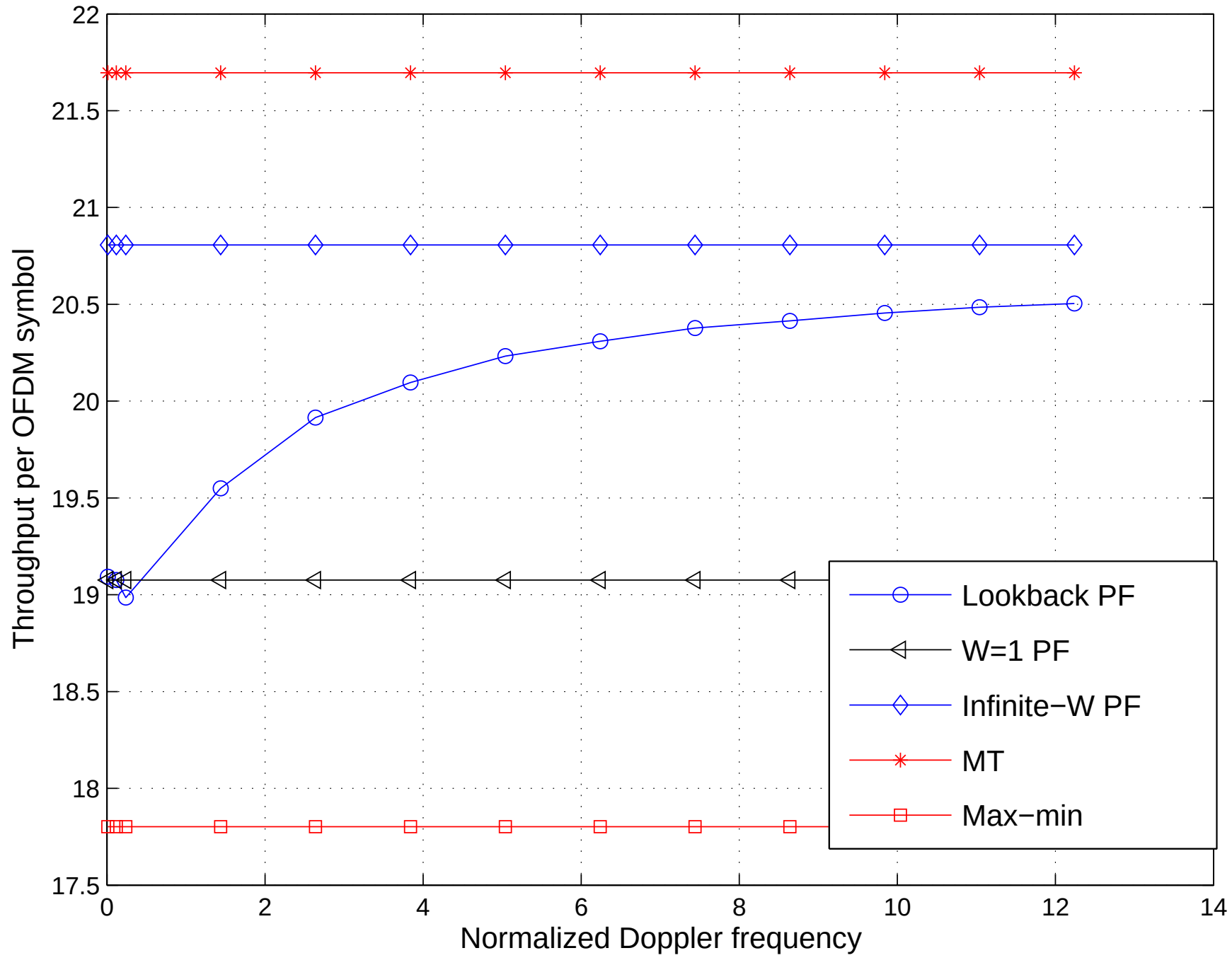

**Fig. 3:** System throughput for systems with inhomogeneous users

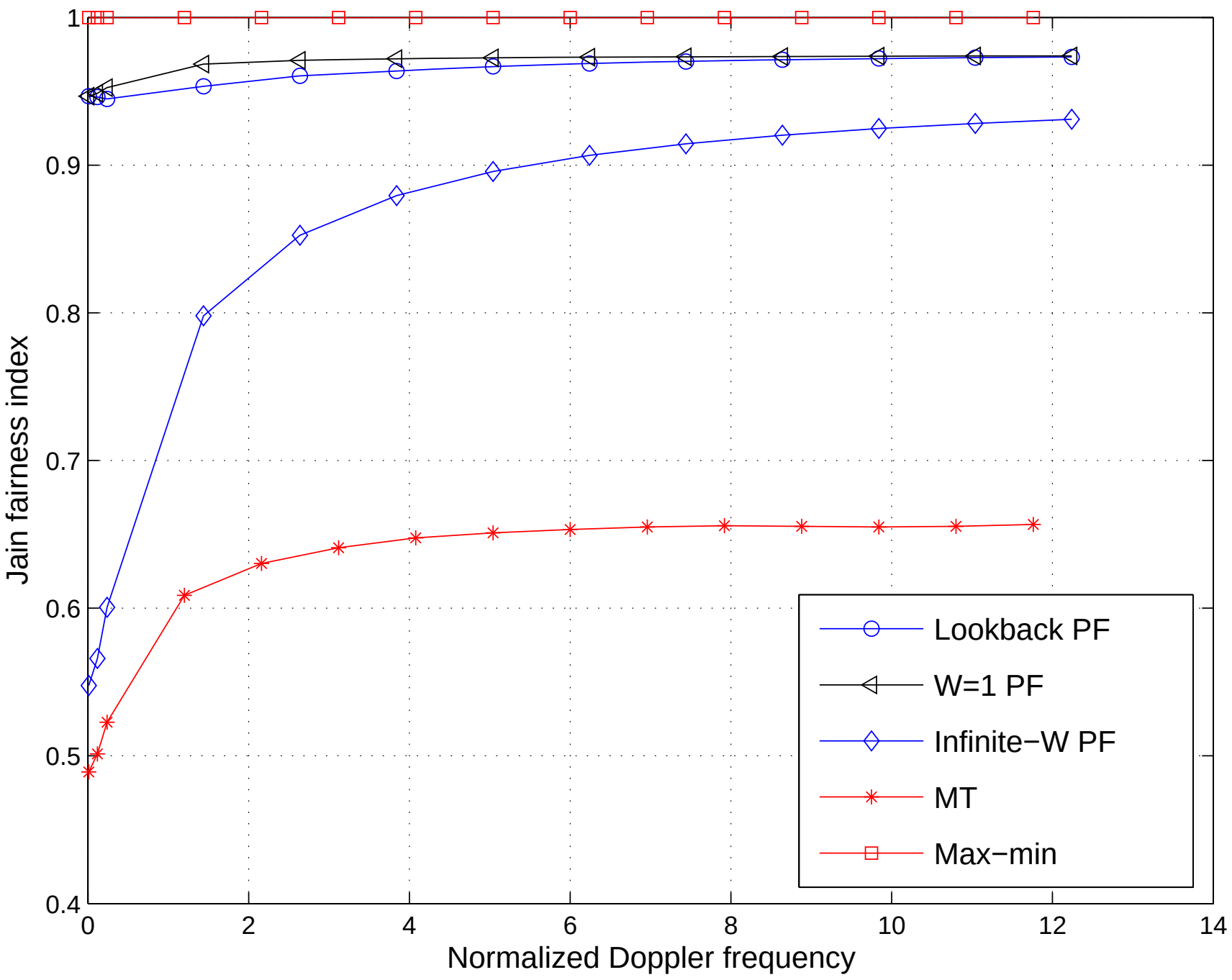

**Fig. 4:** Jain's fairness index for systems with inhomogeneous users

So far, we have investigated the effect of time-domain diversity on the performance of subcarrier assignment schemes. In Fig. 5 and Fig. 6, the effect of channel frequency selectivity is studied. We assume the channels are statistically identical. Specifically, we fix the normalized Doppler frequency to 6 and vary the RMS delay spread of the channel from 0 to 1083*ns*. In particular, 0 RMS delay corresponds to a flat fading channel, while a large RMS delay leads to a highly frequency selective channel, causing $b_{i,k}$ to fluctuate drastically across subcarriers. The mean SNR of all users are equal to 13dB. Fig. 5 shows that although the throughputs of $W$=1 PF and max-min fail to exploit time-domain diversity, they succeed in exploiting the frequency-domain diversity as channel becomes highly frequency selective. When RMS delay is very large, throughput and fairness performance of all schemes converge. This implies that PF, MT, and max-min scheduling schemes all yield the same performance results when there is sufficient diversity in the frequency domain.

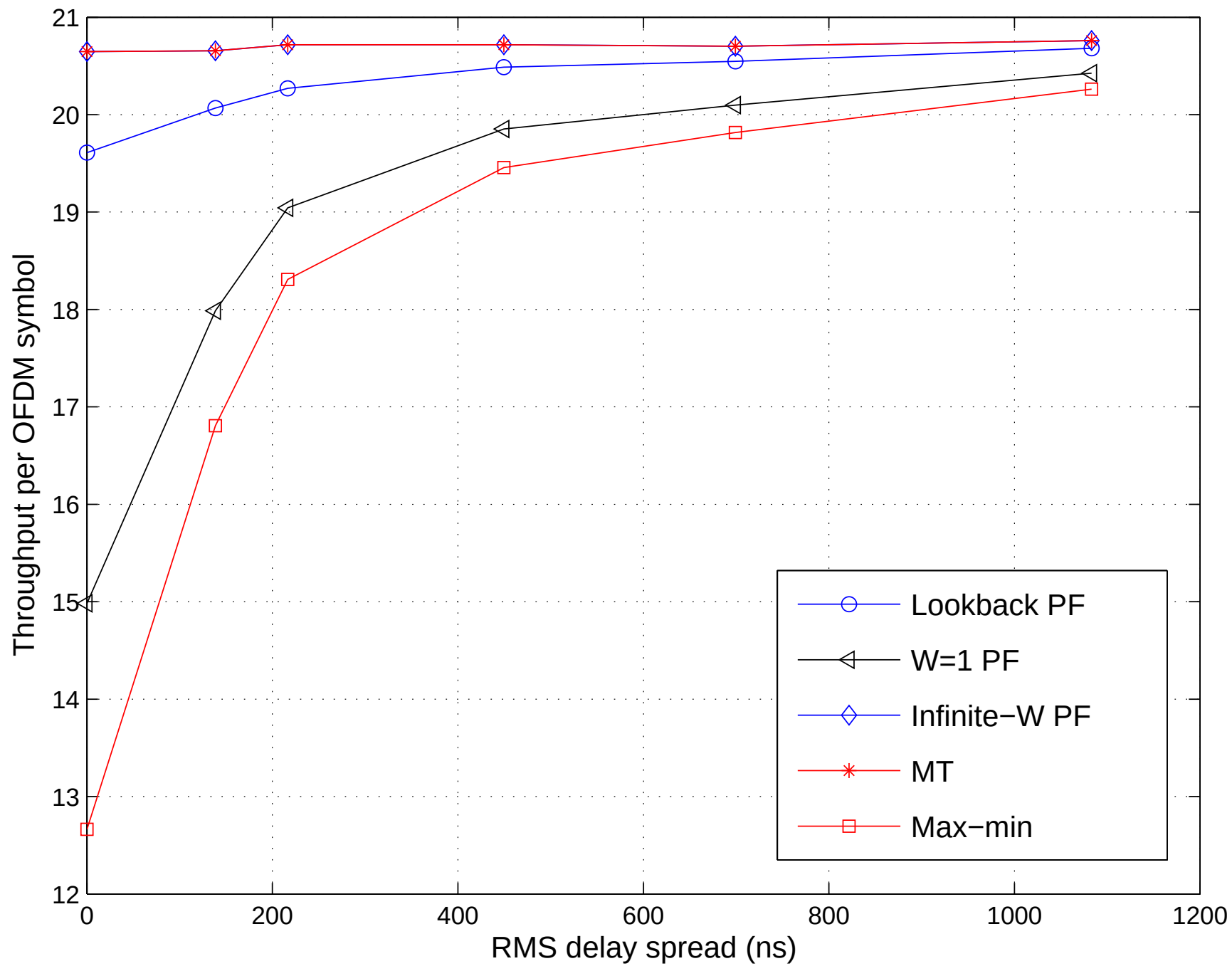

**Fig. 5:** System throughput as a function of RMS delay spread

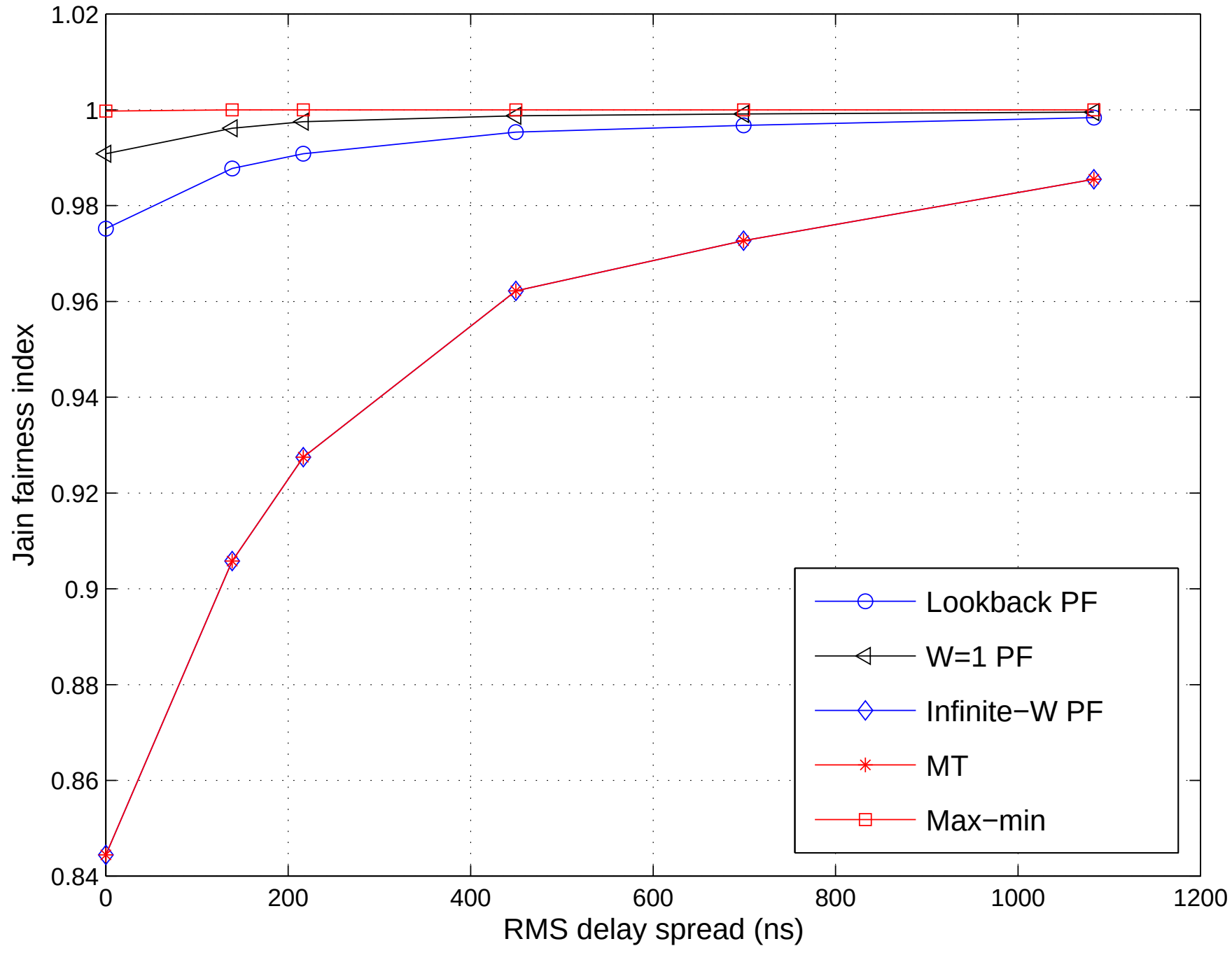

**Fig. 6:** Jain's fairness index as a function of RMS delay spread

We conclude this section with the following observations. PF in general tries to strike a good balance between system throughput and fairness. However, *W*=1 PF, look-back PF, and infinite-*W* PF, do so differently. Look-back PF does so by taking into account the application time window *W*. Blindly fixing *W* to 1 or infinity in the PF scheduler does not take the application requirement into account. For example, fixing *W* to 1 when the application allows for larger *W* may cause the system to under-perform; on the other hand, fixing *W* to infinity when the application requires a small *W*, may be overly aggressive and may cause the underlying application to fail to meet application requirements. Finally, although MT and max-min make good sense in specific situations (e.g., when the different users experience i.i.d channels for MT) they do not perform well in all situations. Look-back PF is an all-around performer in that it automatically takes the different situations into account in the optimization process. Having said that, we remind the reader that the "ensemble average" algorithm of infinite-*W* PF has the advantage that the decision variables $P_{i,k|\mathbf{b}}$ can be pre-computed off-line. It is a convenient low-complexity approximation to the look-back PF scheme when *W* is large.

## V. FURTHER DISCUSSIONS

The analysis and discussion thus far assume the "saturated" or "backlogged" case in which the users always have data to send. In practice, during a session, a user may alternate between active and idle states, and the user may not have any data to send while in idle state. We now discuss how to apply the PF framework to this situation. In addition, we also briefly discuss a plausible approach for the case where different types of applications with different *W* co-exist in the system. For a focus, we only consider the look-back PF scheme with $W < \infty$ here. To limit our scope, we will only sketch the approaches in generalities without going into details. The sketch serves only as a starting point for further investigation.

### *A. PF Scheduling in Non-Saturated Case*

In any time slot *n*, not all users have backlogged data. The problem becomes that of allocating channel airtimes to users with backlogged data. The issue is how to compute $T_i^{(W)}[n]$ among backlogged users. Once we fix the way to compute $T_i^{(W)}[n]$, similar algorithms to those discussed in the previous sections can be used again. We discuss two possibilities here.

The first approach is to simply use the same definition as before. That is,

$$T_i^{(W)}[n] = \frac{1}{\min(n,W)} \sum_{m=\max(1,n-W+1)}^{n} T_i[m] \quad \forall\ i = 1,\ldots,U;\ n \geq 1 \tag{23}$$

However, the optimization is only over the set of users with backlogged data, denoted by $U_b$, as follows:

$$\begin{aligned} &\max y[n] = \sum_{i\in U_b} \log\left( a_i[n-1] + \frac{1}{\min(n,W)} \sum_{k=1}^{S} P_{i,k}[n] b_{i,k}[n] \right) \\ &\text{s.t.} \quad \sum_{i\in U_b} P_{i,k}[n] = 1 \quad \forall\ k = 1,\cdots,S \\ &\qquad P_{i,k}[n] \ge 0 \qquad \forall\ i \in U_b;\ k = 1,\cdots,S \end{aligned} \tag{24}$$

where $a_i[n-1] = \left(\sum_{m=\max(0,n-W+1)}^{n-1} T_i[m]\right) \Big/ \min(n,W)$. The advantage of this approach is its simplicity. It is also appealing from the "fairness" viewpoint in that credit is accumulated to the users who are idle so that they have higher priority when new data arrives.

The second approach is to take each busy period of a user separately in the computation of $T_i^{(W)}[n]$, as follows:

$$T_i^{(W)}[n] = \frac{1}{\min(n - n_i[n] + 1, W)} \sum_{m=\max(n_i[n],n-W+1)}^{n} T_i[m] \quad \forall\ i \in U_b;\ n \ge 1. \tag{25}$$

where $n_i[n]$ is the starting time of the most recent busy period of user *i* looking back from time slot *n*. The corresponding optimization is as follows:

$$\begin{aligned} &\max y[n] = \sum_{i\in U_b} \log\left( a_i[n-1] + \frac{1}{\min(n - n_i[n] + 1, W)} \sum_{k=1}^{S} P_{i,k}[n] b_{i,k}[n] \right) \\ &\text{s.t.} \quad \sum_{i\in U_b} P_{i,k}[n] = 1 \quad \forall\ k = 1,\cdots,S \\ &\qquad P_{i,k}[n] \ge 0 \qquad \forall\ i \in U_b;\ k = 1,\cdots,S \end{aligned} \tag{26}$$

where $a_i[n-1] = \left(\sum_{m=\max(n_i[n]-1,n-W+1)}^{n-1} T_i[m]\right) \Big/ \min(n - n_i[n] + 1, W)$. This approach is more applicable to applications where the delay urgency only starts to "tick" upon the arrival of data to the queue of users. The idle period is not given credit in the airtime allocation here.

With both approaches, there is also the caveat that the allocated airtimes to a particular backlogged user *i* in time slot *n* may not be fully used because of the lack of sufficient backlogged data. In that case, the unused airtimes of user *i* need to be reallocated to the other users. We will not delve into the further details here.

*B. Co-existing User Applications with Different W*

A simple way to take into account applications with different $W$ requirements is to use different $W$ for different users $i$. That is, we simply replace $W$ by $W_i$ in all the previous discussions and equations. In this way, the throughputs of user applications will automatically be smoothed in accordance to their $W_i$ requirements. Note that applications with small $W_i$ will automatically have an advantage over applications with large $W_i$ in the channel airtime allocation process.

## VI. CONCLUSIONS

To conclude, Part II of this paper series has laid down the foundation for PF optimization that takes into account the application time window $W$ when exploiting the time-varying channels for opportunistic airtime allocation. Essentially, in our framework, throughput over $W$ is what the user cares about.

For finite $W$, we have shown how the optimization algorithms in Part I can be adapted for use here through a "look-back" PF optimization formulation. For applications that have very high tolerance for delay, we can assume $W \to \infty$, for which we have formulated a different approach that looks at the ensemble-averaged throughput. The latter formulation has two advantages from the implementation standpoint: (i) Its decision variables can be pre-computed off-line, so that real-time decision making consists of a mere table look-up; (ii) If for each user, the channel statistics of different channels are identically distributed (but not necessarily independently distributed), a reasonable assumption in many practical settings, the multi-channel optimization problem can then be reduced to the single-channel problem. For reasons (i) and (ii), our $W \to \infty$ formulation is a convenient low-complexity approximation to the look-back PF scheme when $W$ is large.

Another interesting aspect to our $W \to \infty$ formulation is that the continuous data-rate case wherein $b_{i,k}$ can adopt a continuum of values is actually easier to solve than the discrete data-rate case wherein $b_{i,k}$ can only adopt finite number of possible values. Specifically, we have developed a set of fixed-point equations for optimality that is amenable to standard fixed-point algorithms.

For numerical studies, this paper has considered cellular OFDM systems. We find that PF scheduling, compared with other scheduling mechanisms, can strike a good balance between system throughput and fairness while taking into account the underlying application time window $W$. We have introduced the concept of a $W$-normalized Doppler frequency and shown that the extent to which the scheduler can exploit opportunistic scheduling is tied to the $W$-normalized Doppler frequency and not just the physical Doppler frequency alone.

This two-part paper series implicitly assumes a centralized implementation of the scheduler or resource allocator. If there are errors in the data collected, the optimization process may indeed not be optimal. The robustness of the algorithms against such errors, and distributed implementations of the algorithms, are interesting subjects for further studies.

REFERENCES:

[1] F. P. Kelly, "Charging and Rate Control for Elastic Traffic" *Euro. Trans. Telecommun.*, vol 8, pp. 7-20, 1997.

[2] P. Viswanath, D. Tse, and R. Laroia, "Opportunistic beamforming using dumb antennas," *IEEE Trans. Inform. Theory*, vol. 48, no. 6, pp. 1277-1294, June 2002.

[3] T. Keller and L. Hanzo, "Adaptive multicarrier modulation: A convenient framework for time-frequency processing in wireless communications," *IEEE Proceedings*, vol. 8, no. 5, pp. 611-640, May 2000.

[4] Y. J. Zhang and K. B. Letaief, "Multiuser Adaptive Subcarrier-and-Bit Allocation with Adaptive Cell Selection for OFDM Systems," *IEEE Trans. Wireless Commun.*, vol. 3 , no. 5, pp. 1566-1575, Sep. 2004.

[5] D. Tse and P. Viswanath, *Fundamentals of Wireless Communication*, Cambridge University Press, May 2005.

[6] L. Jiang and S. C. Liew, "'Proportional Fairness in WLANs and Ad Hoc Networks'," *IEEE Wireless Communications and Network Conference* (*WCNC)*, Mar 2005.

[7] M. S. Bazaraa and C. M. Shetty, *Nonlinear Programming: Theory and Algorithm*, Wiley.

[8] R. Jain, D. Chiu, W. Hawe, "A Quantitative Measure of Fairness and Discrimination for Resource Allocation in Shared Computer Systems", DEC Report, DEC-TR-301, Sept 1984.

[9] TIA/EIA IS-856, "CDMA 2000: High rate packet data air interface specification," Std., Nov. 2000.

[10] P. Bender, P. J. Black, M. Grob, R. Padovani, N. Sindhushyana, and S. Viterbi, "CDMA/HDR: a bandwidth efficient high speed wireless data service for nomadic users," *IEEE Commun. Mag.*, vol. 38, pp. 70-77, July 2000.

[11] E. F. Chaponniere, P. J. Black, J. M. Holtzman, and D. N. C. Tse, "Transmitter directed code division multiple access system using path diversity to equitably maximize throughput," U.S. Patent 6,449,490, Sep. 10, 2002.

[12] H. J. Kushner and P. A. Whiting, "Convergence of proportional-fair sharing algorithms under general conditions," *IEEE Trans. Wireless Commun.*, vol. 3, no. 4, pp. 1250-1259, July 2004.

[13] M. Andrews, "Instability of the proportional fair scheduling algorithm for HDR," *IEEE Trans. Wireless Commun.*, vol. 3, no. 5, pp. 1422-1426, Sep. 2004.

[14] S. Borst, "User-level performance of channel-aware scheduling algorithms in wireless data networks," *IEEE/ACM Trans. Network.*, vol. 13, no. 3, pp. 636-647, June 2005.

[15] H. Kim and Y. Han, "A proportional fair scheduling for multicarrier transmission systems," *IEEE Commun. Lett.*, vol. 9, no. 3, pp. 210-212, March 2005.

## APPENDIX I: Proofs of Theorems 1 and 2

***Proof of Theorem 1:*** For concreteness, the following proof uses the notation of the discrete-rate case. For the continuous-rate case, we can simply replace $p_{\mathbf{B}}(\mathbf{b})$ by $f_{\mathbf{B}}(\mathbf{b})d\mathbf{b}$. Consider a feasible solution $(P_{i,k|\mathbf{b}})$ to the joint-channel problem. The KKT conditions (see Section II of Part I of this paper series) are necessary and sufficient for it to be optimal. Specifically, for each virtual channel $(k,\mathbf{b})$, we require the airtime allocations for any pair of users *i* and *j* to satisfy the following:

1) If $P_{i,k|\mathbf{b}} > 0$ and $P_{j,k|\mathbf{b}} > 0$ then $b_{i,k}\, p_{\mathbf{B}}(\mathbf{b}) / E_{\mathbf{B}}[T_i] = b_{j,k}\, p_{\mathbf{B}}(\mathbf{b}) / E_{\mathbf{B}}[T_j]$ (or equivalently $b_{i,k} / E_{\mathbf{B}}[T_i] = b_{j,k} / E_{\mathbf{B}}[T_j]$)
2) If $P_{i,k|\mathbf{b}} > 0$ and $P_{j,k|\mathbf{b}} = 0$ then $b_{i,k}\, p_{\mathbf{B}}(\mathbf{b}) / E_{\mathbf{B}}[T_i] \ge b_{j,k}\, p_{\mathbf{B}}(\mathbf{b}) / E_{\mathbf{B}}[T_j]$ (or equivalently $b_{i,k} / E_{\mathbf{B}}[T_i] \ge b_{j,k} / E_{\mathbf{B}}[T_j]$).

Consider the single-channel optimization problem (9) on a channel *k*. Let $\mathbf{B}_k$ and $\mathbf{b}_k$ denote the *k*th column of $\mathbf{B}$ and $\mathbf{b}$, respectively. Suppose that $(P^*_{i,k|\mathbf{b}_k})$ is an optimal solution to the single-channel problem on channel *k*, and it yields the expected throughput $E_{\mathbf{B}_k}[T^*_{i,k}]$ for user *i* on channel *k*. If we combine the solutions of all channels, the corresponding solution to the joint-channel gives $E_{\mathbf{B}}[T_i] = \sum_k E_{\mathbf{B}_k}[T^*_{i,k}] = S \cdot E_{\mathbf{B}_k}[T^*_{i,k}]\ \forall\ i$. Let the combined solution be $(P_{i,k|\mathbf{b}})$. Consider the virtual channel $(k,\mathbf{b})$. The nature of the combined solution is such that $P_{i,k|\mathbf{b}} = P^*_{i,k|\mathbf{b}_k}$ for virtual channel $(k,\mathbf{b})$. We need to show that the combined solution satisfies KKT conditions 1 and 2 above, and therefore it is optimal with respect to the joint-channel optimization problem also.

Consider condition 1. If $P_{i,k|\mathbf{b}} > 0$ and $P_{j,k|\mathbf{b}} > 0$, then $P^*_{i,k|\mathbf{b}_k} > 0$ and $P^*_{j,k|\mathbf{b}_k} > 0$. Since $(P^*_{i,k|\mathbf{b}_k})$ is an optimal solution for the single-channel case, the KKT condition for the single-channel case, $b_{i,k} / E_{\mathbf{B}_k}[T^*_{i,k}] = b_{j,k} / E_{\mathbf{B}_k}[T^*_{j,k}]$, must be satisfied. From the previous paragraph $E_{\mathbf{B}_k}[T^*_{i,k}] = E_{\mathbf{B}}[T_i]/S$ and $E_{\mathbf{B}_k}[T^*_{j,k}] = E_{\mathbf{B}}[T_j]/S$. Therefore, $b_{i,k} / E_{\mathbf{B}}[T_i] = b_{j,k} / E_{\mathbf{B}}[T_j]$. Thus, KKT condition 1 is satisfied for the combined solution also. Similar argument applies for KKT condition 2. □

***Proof of Theorem 2*****:** The proof is similar to the proof of Theorem 1. Consider the KKT conditions 1 and 2 in the proof of Theorem 1. If we adopt the strategy of maximizing system throughput, then $E_{\mathbf{B}}[T_i]$ for different $i$ are equal because of the i.i.d assumption. In addition, for user $i$, $P_{i,k|\mathbf{b}} > 0$ only if $b_{i,k} = \max_j b_{j,k}$. The satisfaction of KKT conditions 1 and 2 is then obvious. □